\documentclass[twocolumn,prb,superscriptaddress,floatfix]{revtex4}
\usepackage{graphicx}
\usepackage{setspace}
\usepackage{color}

\newcommand {\be} {\begin{eqnarray}}
\newcommand {\ee} {\end{eqnarray}}

\newcommand{\degree}{\ensuremath{^\circ}}

\begin{document}

\title{Dilatometric Study of LiHoF$_{4}$ In a Transverse Magnetic Field}
\author{J.L. Dunn}
\affiliation{GWPI, University of Waterloo}
\affiliation{Department of Physics and Astronomy, University of Waterloo, Waterloo, Ontario, Canada, N2L 3G1}
\author{C. Stahl}
\affiliation{Universit\"at Stuttgart, Universitätsbereich Stadtmitte, Postfach 10 60 37, 70049 Stuttgart}
\author{Y. Reshitnyk}
\affiliation{Department of Physics and Astronomy, University of Waterloo, Waterloo, Ontario, Canada, N2L 3G1}
\author{W. Sim}
\affiliation{Department of Physics and Astronomy, University of Waterloo, Waterloo, Ontario, Canada, N2L 3G1}
\author{R.W. Hill}
\affiliation{GWPI, University of Waterloo}
\affiliation{Department of Physics and Astronomy, University of Waterloo, Waterloo, Ontario, Canada, N2L 3G1}

\date{\today}

\begin{abstract}
Theoretical and experimental work have not provided a consistent picture
of the phase diagram of the nearly ideal Ising ferromagnet LiHoF$_{4}$ in a
transverse magnetic field. Using a newly fabricated capacitive
dilatometer, we have investigated the thermal expansion and
magnetostriction of LiHoF$_{4}$ in magnetic fields applied perpendicular to the
Ising direction. Critical points for the ferromagnetic phase transition
have been determined from both methods in the classical paramagnetic to
ferromagnetic regime.  Excellent agreement has been found with existing
experimental data suggesting that, in this regime, the current theoretical
calculations have not entirely captured the physics of this interesting
model system.
\end{abstract}

\maketitle

\section{\label{intro}INTRODUCTION}

The study of ideal Ising systems has attracted considerable attention, owing to the computational tractability of theoretical models and the discovery of some materials which exhibit nearly ideal Ising characteristics\cite{wolffBJP2000}. One such material is LiHoF$_{4}$, which becomes ferromagnetic at the Curie temperature, $T_c=$ 1.53 K\cite{cookeprc1975, nikkelprb64}. The magnetic behavior is due the spin on the Ho$^{3+}$ ion.  The Ising anisotropy is provided by the crystalline electric field which splits the multiplet of the Ho$^{3+}$ states yielding a groundstate doublet with the spins pointing parallel or antiparallel to the [001] axis of the crystal.  The first excited state is separated from the groundstate by approximately 11 K \cite{hansen}.  As the temperature is lowered the magnetic ordering proceeds via a second-order phase transition.  In principle, this is a well-understood classical phenomenon driven by thermal fluctuations\cite{sachdevqpt}.  In this case, the system dimension is equal to the critical dimension leading to measurable logarithmic corrections to mean-field critical exponents\cite{nikkelprb64}.

\begin{figure}
\begin{center}
\scalebox{0.65}{\includegraphics{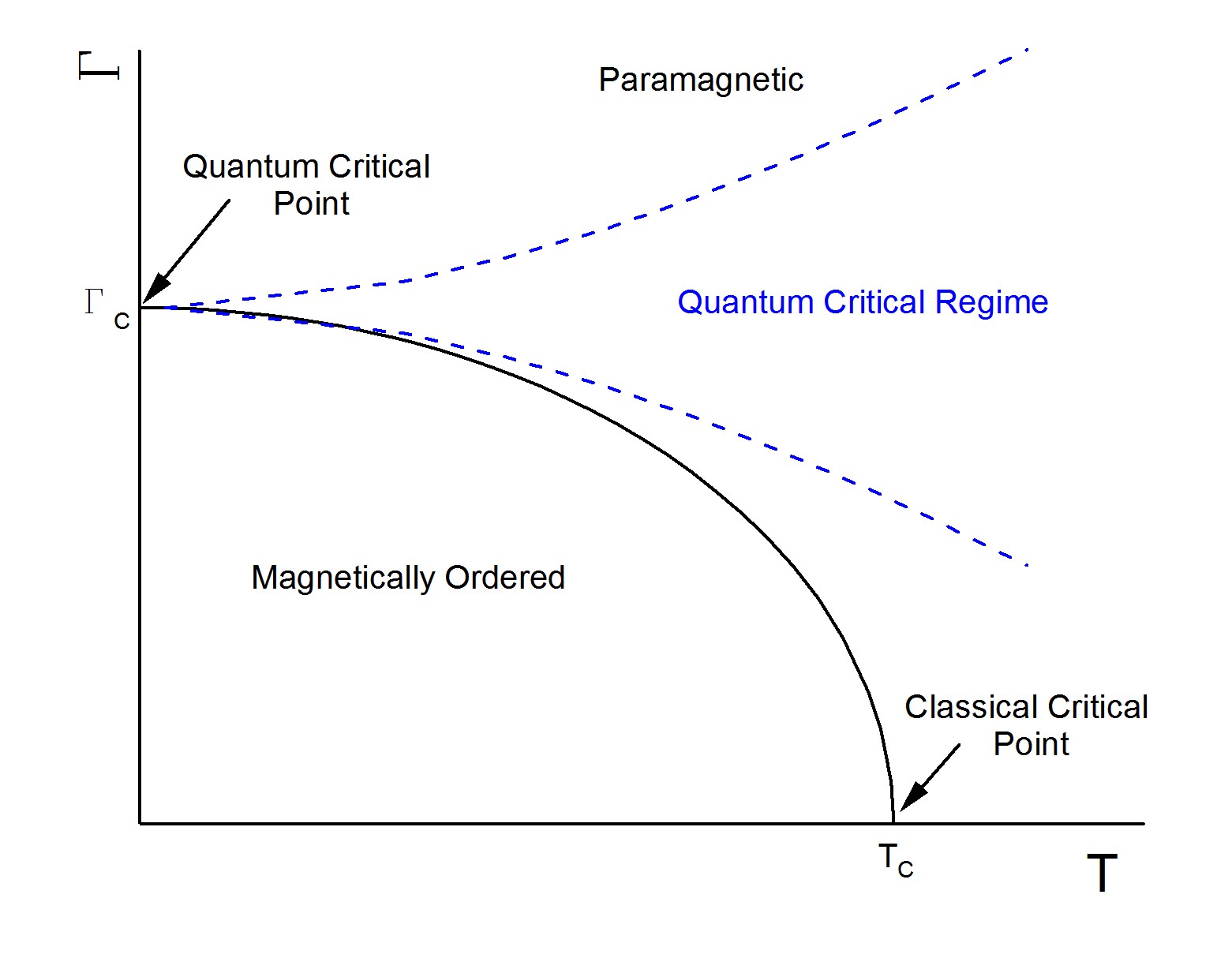}} 
\caption[Schematic of a phase diagram exhibiting quantum critical behavior.]{\label{fig:PD1st}A schematic phase diagram of the transverse field Ising model where $T$ is the temperature and $\Gamma$ is the effective transverse field parameter.  The solid line separates the magnetically ordered phase from the paramagnetic phase.}
\end{center}
\end{figure}

The application of a magnetic field transverse to the Ising (easy) axis of LiHoF$_{4}$ splits the ground state doublet, and introduces quantum mechanical tunneling between the Ising states of the Ho$^{3+}$ ions in the presence of the crystalline electric field. The associated quantum fluctuations lead to a suppression of the Curie temperature with increasing field\cite{bitko}, culminating in a quantum phase transition when the transition temperature is suppressed to zero.    This physics is thought to be captured by one of the simplest theoretical models to describe a quantum phase transition, namely the transverse-field Ising model. Originally proposed by de Gennes\cite{deGennes},  the Hamiltonian for this model is given by\cite{tabeiprb78, schecterprb78}:

\be
\label{eq:IsingHam}
H= -\frac{1}{2} \sum_{i>j}J_{ij} \sigma_{i}^{z}\sigma_{j}^{z} + \Gamma \sum_{i}\sigma_{i}^{x}
\ee

Here $J_{ij}$ are coupling constants between spins, $\sigma_{i}^{x,y,z}$ are Pauli operators, and $\Gamma$ is the effective transverse field parameter that represents the mixing of the two Ising states in the presence of the physical transverse magnetic field. For $\Gamma\neq0$, quantum mechanical fluctuations in the Ising order occur even at zero temperature.  Increasing the fluctuations by increasing the transverse field eventually leads to a disordering of the magnetically ordered phase at a critical value of $\Gamma = \Gamma_c$.  This is the quantum critical point. At finite temperature, the quantum fluctuations lead to a suppression of the transition temperature compared with the zero-transverse-field value.  A phase diagram is shown schematically in Fig. \ref{fig:PD1st}.

Such quantum phase transitions have attracted considerable attention in the physics community recently \cite{RMP-sondhi}. In this respect, the transverse field Ising model provides one of the simplest realisations of such fascinating and topical physics. Paradoxically, for the application of the transverse field Ising model to LiHoF$_4$, whilst the quantum critical regime appears moderately well-understood \cite{chakrabortyprb70, ronnowprb75},  a significant discrepancy exists between the experimental data and the theoretical description of the phase diagram  in the predominantly classical regime close to the zero-field transition temperature\cite{chakrabortyprb70, ronnowprb75, tabeiprb78}. It is in this regime where a complete description requires a full understanding of the effects of temperature on the quantum fluctuations.  Ultimately, this may prove to be the most relevant physics for understanding the novel behaviour related to quantum fluctuations in real systems occurring at finite temperatures\cite{Coleman}.

Using measurements of magnetic susceptibility, Bitko \textit{et al.} \cite{bitko} were the first to report the phase line of the ferromagnetic state in LiHoF$_4$ in a transverse magnetic field using measurements of magnetic susceptibility.  A reasonable theoretical fit to the data was presented using mean-field theory albeit with two adjustable parameters; the Land\'{e} g-factor and effective dipole coupling strength to characterise the effect of magnetic field and temperature, respectively.

Recognizing the limitations of such a simplified theoretical approach,  Chakraborty \emph{et al.} \cite{chakrabortyprb70} developed a full microscopic Hamiltonian that purportedly captured all the relevant physics of this transverse field Ising system, notably the effects of quantum fluctuations and the domain structure of the ferromagnetic state.  Parameters in the theory were fixed where possible by results from spectroscopic\cite{christensenprb1979} and susceptibility\cite{hansen} measurements leaving one adjustable parameter as the antiferromagnetic Heisenberg exchange coupling.  This quantity was introduced to explain the measured zero-field value of the ferromagnetic ordering temperature which is lower by some 25\% over the quantum Monte Carlo value fixed by the known parameters\cite{chakrabortyprb70}.  Using quantum Monte Carlo numerical techniques, a phase line was computed that agreed well with the data particularly in the low temperature limit where quantum fluctuations are significant.    Nonetheless, a discrepancy with the data close to the zero-field transition was apparent and reported as reflecting the potential uncertainties in the crystal- field parameters for LiHoF$_{4}$.

Further theoretical work on the transverse field Ising model as it relates to LiHoF$_{4}$ followed by R\o nnow \emph{et al.} \cite{ronnowprb75}.  This study addressed neutron scattering spectra \cite{ronnowsci308} as well as the phase line measured earlier \cite{bitko}. The critical points determined from neutron scattering experiments lacked close agreement at low transverse field compared to previous susceptibility work \cite{bitko}.  For the theoretical calculation, the crystal-field parameters were derived from spectroscopic measurements\cite{salaunjpcm1997, magarinoprb1980, christensenprb1979} on Ho-doped LiYF$_{4}$ which agreed closely with previously determined values\cite{hansen}.  As in the Chakraborty work \cite{chakrabortyprb70}, the computed phase line overestimated the rate of suppression of the ferromagnetic state at very low transverse fields.  However, given that now different crystal-field parameters have led to a similar discrepancy between theory and experiment as had been seen earlier, it is unlikely that this is the origin.

Most recently, a substantial theoretical effort by Tabei \emph{et al.} \cite{tabeiprb78} has been reported that, in particular, attempts to resolve the differences between the theoretical phase line and experiment at low transverse fields\cite{chakrabortyprb70, ronnowprb75}.  Using the fact that in this regime the quantum fluctuations are small, they were introduced perturbatively into a classical Hamiltonian.  The calculation then proceeded using classical Monte Carlo techniques that were both significantly different and simpler to implement than their quantum Monte Carlo counterparts.  The possible sources of discrepancy were separated into ones which were computational in nature and ones that resulted from inadequacies in the model Hamiltonian.  The authors concluded that the discrepancy was not of computational origin, nor did it stem from uncertainties in crystal-field parameters.  Consequently, it was most likely due to shortcomings of the model Hamiltonian (Eq. \ref{eq:IsingHam}) used. Furthermore, they also made the suggestion that the experimental determination of the phase-diagram should be revisited.  This provided substantial motivation for the study reported here.

As outlined above, there have been several attempts to improve the theoretical understanding of the phase diagram of LiHoF$_{4}$, while little has been done to repeat the experimental measurement of the phase diagram.  Here we report a detailed experimental study of the phase line using thermal expansion and magnetostriction measurements. We focus exclusively on the high temperature -- low field regime where there are markedly few data points from the original experimental investigation\cite{bitko} and where the theory is most challenged.
Our results show good agreement with the existing experimental data indicating that there is still more theoretical work to be done. The rest of this article will briefly discuss our experimental technique, the methods used to determine critical points herein, and then present the results from this investigation into the transverse field phase diagram of LiHoF$_{4}$.

\section{\label{apparatus}APPARATUS and EXPERIMENTAL METHOD}
Previous experimental investigations have made use of magnetic susceptibility\cite{bitko}, neutron scattering\cite{ronnowsci308} and specific heat \cite{nikkelprb64} (zero-field only) to determine critical points in the phase diagram of LiHoF$_{4}$. For the study reported here, capacitive dilatometry has been used to measure the thermal expansion and magnetostriction of a LiHoF$_{4}$ sample. A detailed description of this device will be limited in this article, and we refer interested readers to a future publication\cite{Dunn}.

The LiHoF$_{4}$ sample is taken from a commercially produced single crystal\cite{LiHoF4}.  It is shaped as a roughly semi-circular plate, where the easy axis [001] of LiHoF$_{4}$ is oriented approximately 25$^{\circ}$ from the
flat edge of the semi-circle. The sample is approximately 1 mm thick, 5.5 mm tall,
and about 3.5 mm at it's widest point. The easy [001] axis of the LiHoF$_{4}$ sample was
determined to within $\pm 1 \degree$ using a commercial Lau\'e diffraction
setup, and the ends of the crystal subsequently polished to be orthogonal to this easy axis
(within $\pm 5 \degree$). When placed in the dilatometer, length changes
along the [001] axis were measured as temperature and magnetic field were varied.

The dilatometer used here is conceptually similar to that reported by Schmiedeshoff \textit{et al.}
\cite{schmiedeshoff}. It is
fabricated from 99.995\% pure silver and differs in that the moveable
capacitor plate is mounted to a BeCu spring known for its flexibility at low temperatures. The sample is then pressed
against this spring, and does not make direct contact with the moveable
capacitor plate. Samples are thermally linked to a sample mount (made of
99.995\% pure silver) using a silver paint\cite{Dupont-AgPaint}. An uncalibrated Cernox CX-1030
resistance thermometer is also fastened directly to this mount and is used to verify the thermal
stability of the sample.

A pumped $^{4}$He cryostat is used to cool the sample mounted in the
dilatometer to approximately 1~K.  The temperature is controlled using a Lakeshore
331 Temperature controller and measured by a calibrated Cernox CX-1030
thermistor located on the cold stage of the pumped $^{4}$He cryostat. An
experimental trial consists of a temperature sweep through the region of
interest ($\approx$~1.2~K to 1.8~K) at a rate of 7.5~mK/min.

Capacitance was measured using a commercial Andeen-Hagerling 2500A
capacitance bridge, with a resolution of $10^{-7}$ pF,
which equates to a dilation of 0.002~\AA~ at 17 pF. The
measured capacitance is converted to capacitor plate separation
using an expression for a tilted plate capacitor\cite{swenson98}:
\be
\label{eq:tiltcap}
D = \frac{\epsilon_{0} A}{C} \Big{[} 1 + \Big{(} \frac{C}{C_{\textrm{\tiny max}}}
\Big{)}^{2} \Big{]}
\ee
Where $\epsilon_{0}$ is the permittivity of free space, $A$ is the effective
capacitor plate area, $C_{\textrm{\tiny max}}$ is the maximum measured capacitance before
the capacitor plates short, and $C$ is the measured capacitance. Prior to
beginning a measurement, the maximum capacitance (C$_{\textrm{\tiny max}}$, typically
$\approx$~50~pF) is measured.  This  is the measured capacitance
just before the plates touch and electrically short (leading to $C=0$~pF), and
provides enough information to take into account the tilt of the capacitor
plates. Once typical operating conditions are taken into account, length
changes greater than 0.2~\AA~ can be resolved. A measured change in the capacitor plate spacing is directly related to a change in the sample length, $dD = -dL$.  However, the measured length changes will include the effect of the dilatometer as
well as the LiHoF$_{4}$ sample. To decouple the behavior of the sample from
the measured capacitance changes two pieces of information are required: the
behavior of the dilatometer with a silver sample (known as the cell
effect), and the behavior of the dilatometer with LiHoF$_{4}$. Ultimately, the quantity of interest is the thermal expansion coefficient, $\alpha(T)$, which is the normalised rate of change of length with temperature.  The relation
between the measured thermal expansion coefficient, and the thermal expansion coefficient of
LiHoF$_{4}$ is\cite{schmiedeshoff}:
\be
\label{eq:thermexp}
\alpha_{\text{\tiny LiHoF}_{4}} = \frac{1}{L} \frac{dL}{dT} \bigg |_{\textrm{\tiny Cell+LiHoF}_{4}} -
\frac{1}{L} \frac{dL}{dT} \bigg |_{\textrm{\tiny Cell+Ag}} + \alpha_{\text{\tiny Ag}}
\ee
where $\alpha_{\text{\tiny Ag}}$ is the thermal expansion coefficient of silver\cite{alpha-silver}.

\section{\label{results}RESULTS \& DISCUSSION}
\subsection{\label{te}Thermal Expansion}

\begin{figure}
\begin{center}
\scalebox{0.65}{\includegraphics{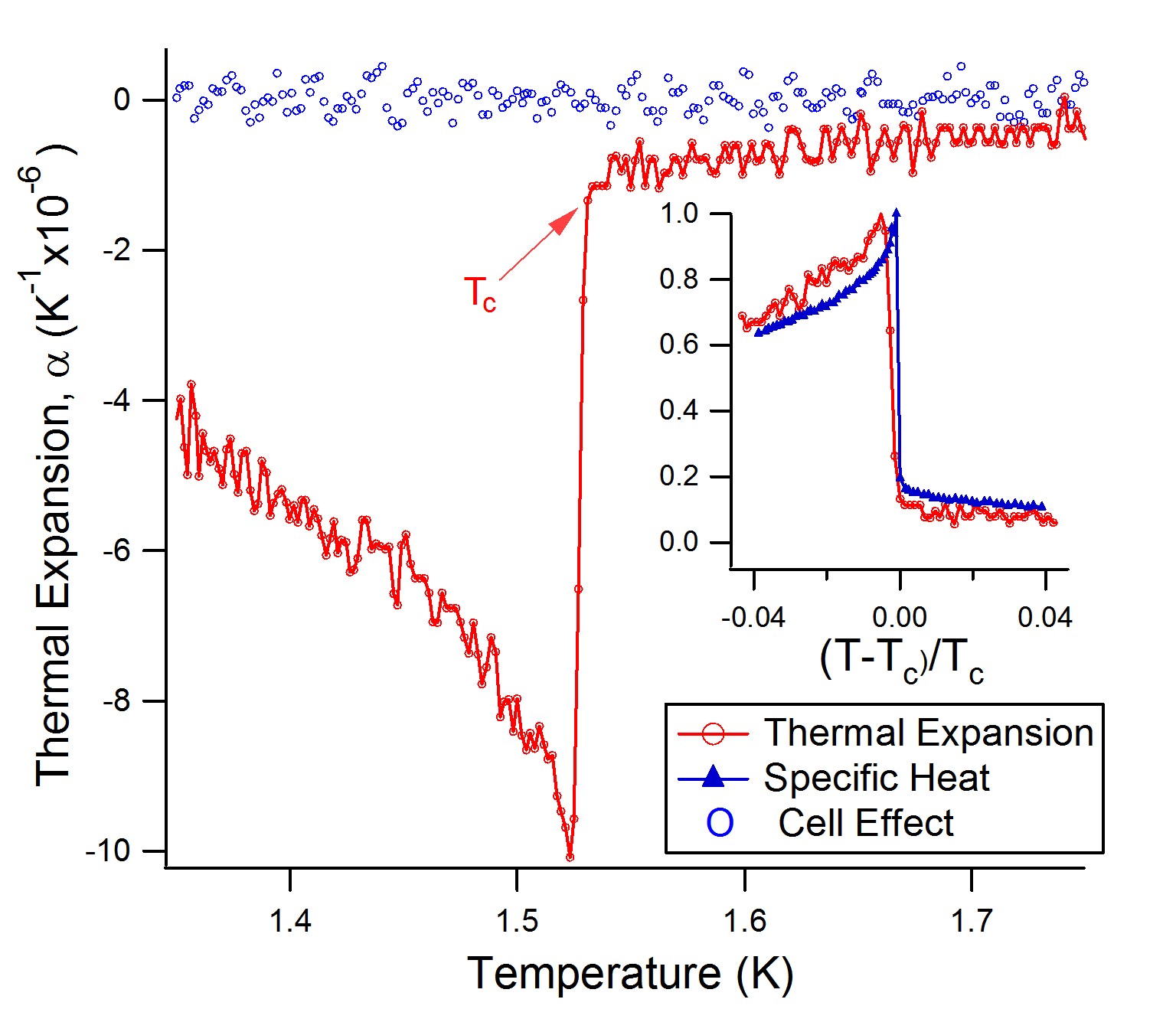}} 
\caption[Zero-field measurement of $\alpha$ of LiHoF$_{4}$ compared to specific heat measurement of Ref. \onlinecite{nikkelprb64} and cell effect.]{\label{fig:0T}Zero-field measurement of the thermal expansion coefficient, $\alpha$, of LiHoF$_{4}$ and the cell effect as a function of temperature.  The critical temperature, $T_c$ is indicated, being the first deviation from the paramagnetic state. Inset: Normalised thermal expansion coefficient and specific heat measurements \cite{nikkelprb64} as a function of reduced temperature. }
\end{center}
\end{figure}

\emph{Zero field}: In Fig. \ref{fig:0T}, the thermal expansion of LiHoF$_4$ is plotted in the absence of any magnetic field. The transition from the paramagnetic to the ferromagnetic state as the temperature is lowered results in a sudden jump in the thermal expansion coefficient.  Consistent with previous experimental\cite{bitko} and theoretical work\cite{chakrabortyprb70}, the critical temperature, $T_c$, is defined as the first deviation from the behaviour in the paramagnetic state as the temperature is decreased. This point is determined from the intersection of a polynomial fit to the temperature dependence of the thermal expansion above the transition and a linear fit to the steepest section of the thermal expansion during the transition (increasing the order of either fit made no appreciable difference). The error in this determination was estimated by varying the range of the fitted curves near the critical region by up to $\pm$0.05~K (10\% of the temperature range), and observing the range of $T_c$'s that result.

Using the above analysis, the transition temperature in zero field is $T_c = 1.532 \pm 0.005$ K.  In the inset to Fig.~\ref{fig:0T}, we plot the thermal expansion data and high-resolution specific heat data \cite{nikkelprb64}.
One can relate the thermal expansion coefficient ($\alpha$) to the specific heat ($C_{v}$) through the Gr\"uneisen parameter ($\gamma$) and the bulk modulus ($B$): $\alpha(T) = (\gamma/3B)~ C_{v}(T)$. Assuming the Gr\"uneisen parameter and the bulk modulus are temperature independent, it is apparent that the discontinuities observed in specific heat measurements should be reproduced in a thermal expansion measurement.  To aid close comparison, the thermal expansion data is inverted and both measurements are normalised to their peak values while being plotted against reduced temperature with respect to their individual transition temperatures.  The qualitative agreement is excellent (as seen inset in Fig.~\ref{fig:0T}).  Quantitatively, there is a smaller discrepancy between the absolute value of transition temperatures ($1.532 \pm 0.005$ K and $1.5384$ K) while the transition width is larger by 4.4~mK in the thermal expansion measurements. Both effects can reasonably be attributed to the swept temperature method that was used in this experiment.

\begin{figure}
\begin{center}
\scalebox{0.65}{\includegraphics{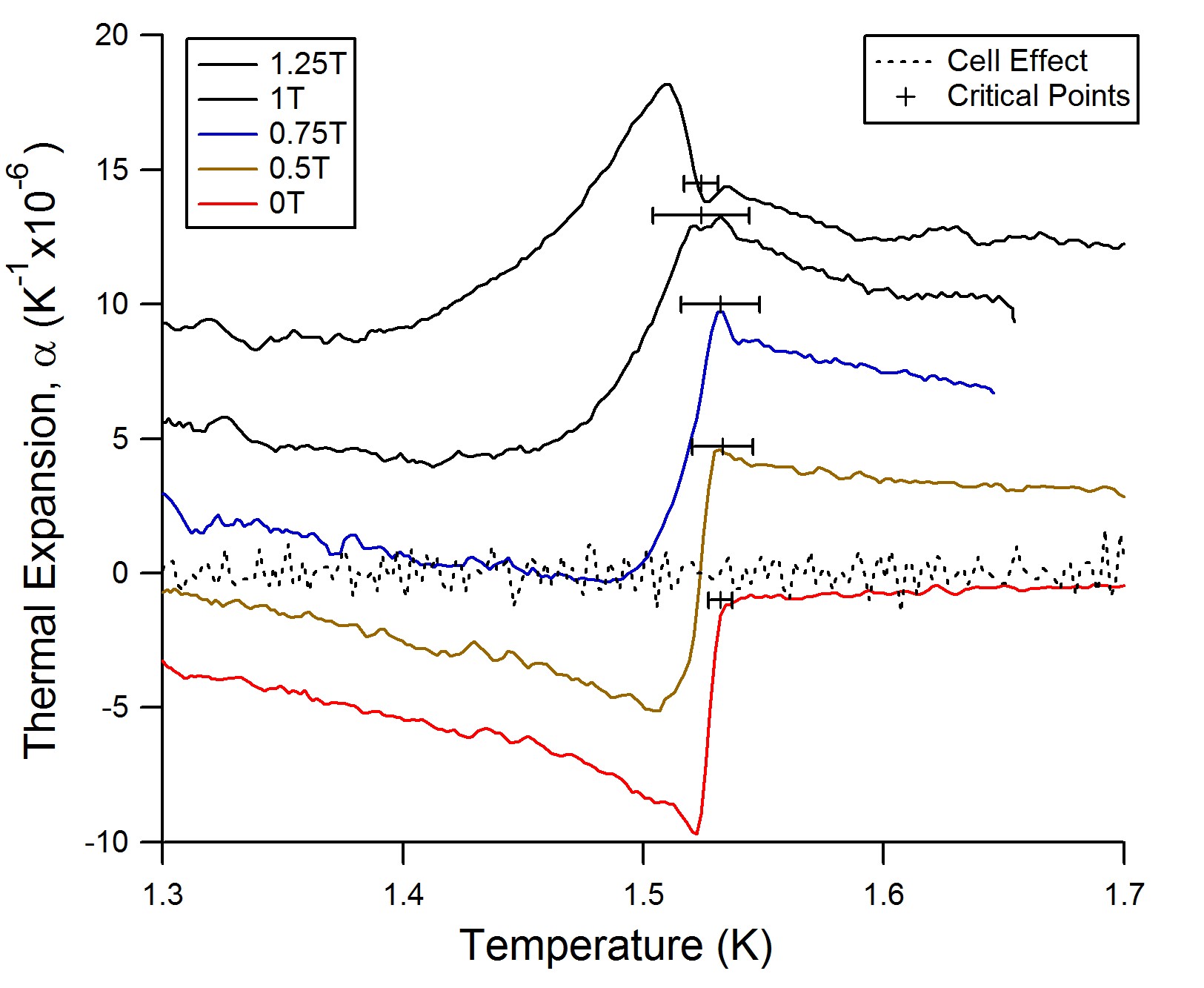}} 
\caption[Low-field measurement of $\alpha$ for LiHoF$_{4}$.]{\label{fig:lowT}Measurement of the thermal expansion coefficient, $\alpha$, for LiHoF$_{4}$ up to an applied transverse field of 1.25~T. The critical temperature at each field is indicated.  The curves at different field strengths are offset by $3\cdot 10^{-6}$ per curve.}
\end{center}
\end{figure}

\begin{figure}
\begin{center}
\scalebox{0.65}{\includegraphics{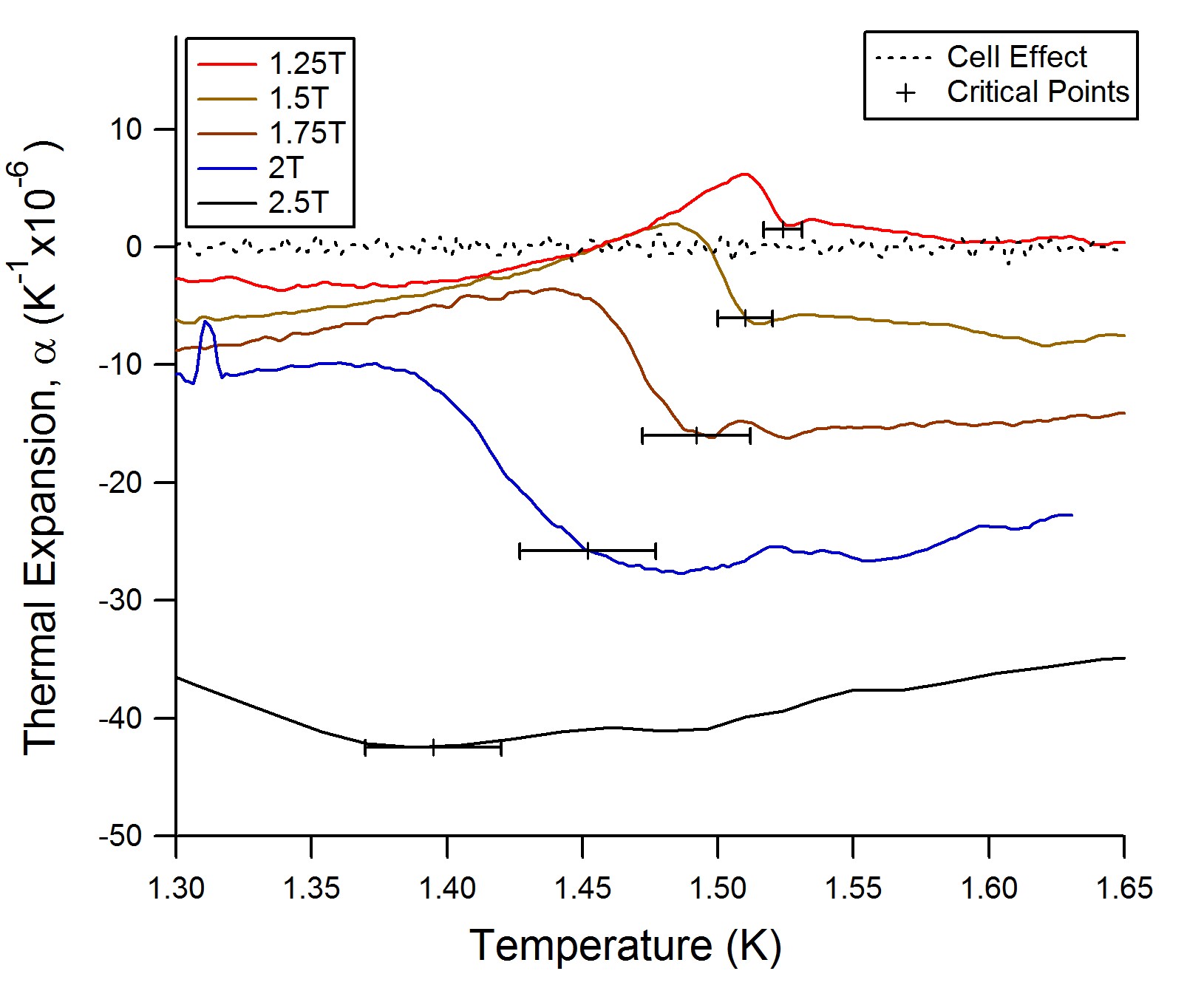}} 
\caption[High-field measurement of $\alpha$ for LiHoF$_{4}$.]{\label{fig:highT}Measurement of the thermal expansion coefficient, $\alpha$, for LiHoF$_{4}$, from 1.25~T to 2.75~T. The critical temperature at each field is indicated.  The curves at different fields are offset by $-5\cdot 10^{-6}$ per curve.}
\end{center}
\end{figure}

\emph{Transverse Magnetic Field}:  The evolution of the temperature dependence of the thermal expansion in transverse field is shown in Figs. \ref{fig:lowT} and \ref{fig:highT}.  In these figures the curves at each field have been offset for clarity.  Three qualitative features are apparent in this data set.  First, the temperature dependence of the thermal expansion coefficient in the paramagnetic state evolves from low positive gradient towards a negative gradient of slightly larger magnitude at a field of 1 T and then returns to a positive gradient by a field of 2 T.  Second, in the ferromagnetic phase, the temperature dependence transitions from concave downwards in zero field and low fields to concave upwards at fields above 1 T.  Third, the transition region itself is modified to accommodate these variations, starting as a step downwards upon entering the ferromagnetic phase and evolving to being a step upwards for fields of 1 T and above.  This is accompanied by a significant broadening of the transition region as the field is increased.

As in the zero field measurement, the transition temperature is defined as the first deviation from the monotonic behaviour that is considered the paramagnetic phase.  The same analysis as described above is used to extract a transition temperature at each magnetic field.  As the transition width broadens with increased magnetic field, the uncertainty in the transition temperature increases and assumes a larger error bar.

\subsection{\label{ms}Magnetostriction}
\begin{figure}
\begin{center}
\scalebox{0.65}{\includegraphics{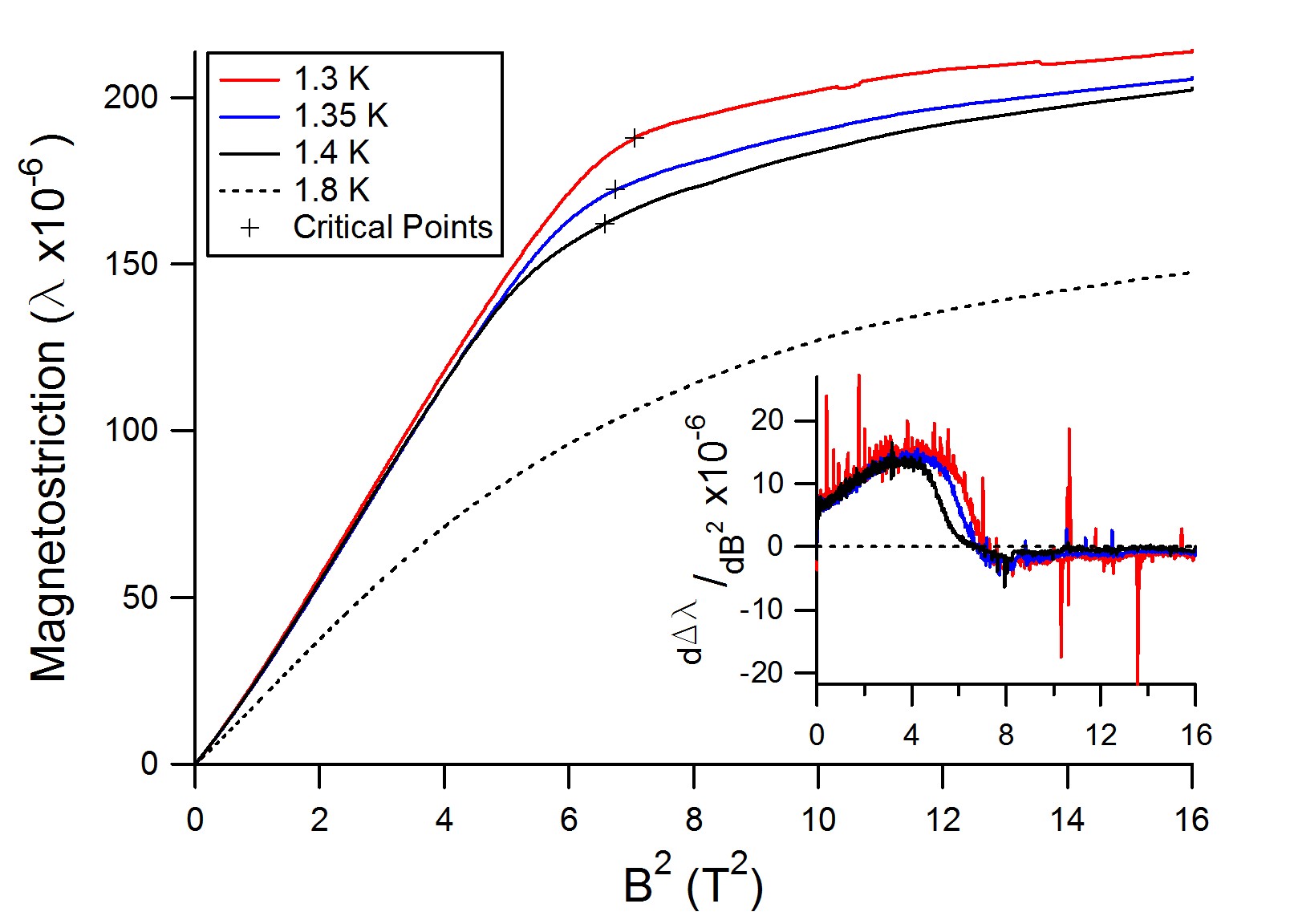}} 
\caption[Magnetostriction measurements of LiHoF$_{4}$.]{\label{fig:MS}Magnetostriction, $\lambda = \frac{\Delta L}{L}$, as a function of transverse field squared, B$^{2}$) measurements performed on LiHoF$_{4}$, in the minimum demagnetization orientation. Inset: $\frac{d\Delta\lambda}{dB^{2}}$ as a function of $B^{2}$. The critical field is the determined when the derivative equals zero.}
\end{center}
\end{figure}

Magnetostriction, $\lambda = \Delta L/L$, is the change in length of a material in response to a magnetic field.
The magnetostriction of LiHoF$_4$ was measured at $T =$ 1.3, 1.35, 1.4 and 1.8~K and the results are plotted in Fig.~\ref{fig:MS} as a function of $B^2$.   Since the measurement at $T = 1.8$~K is entirely in the paramagnetic region, by subtracting this from the reponse at the lower temperatures, $\Delta \lambda = \lambda (T) - \lambda (1.8 \text{K})$, we expose the magnetic response more clearly.  The inset shows the first derivative of $\Delta \lambda$ at 1.3, 1.35 and 1.4~K with respect to the squared transverse field. While the raw magnetostriction signal does not exhibit any distinct features during the phase transition as seen in the thermal expansion, there is a distinct change in this derivative  which we interpret as the field at which ferromagnetic state is entered as the field is lowered.  The critical field is therefore determined to be the point at which the first derivative equals zero.  The uncertainty associated with this value is dominated by the uncertainty of the magnetoresistance of the thermistor used to control the temperature.

\subsection{\label{demag}Demagnetization Effects}
The irregular shape of the sample means that the demagnetization factor due to uncompensated spins on the surface of the material is difficult to calculate.  However,  we can assess the impact of the demagnetization effect on our results by orienting either the semi-circular face or the thin edge of the crystal perpendicular to the transverse field.  These two experimental orientations represent a maximum and minimum demagnetization factor for this particular sample.  In the following we estimate the difference in the magnitude of the demagnetization factor between the two orientations.

The demagnetization correction can be written as $H_{\text{\tiny Internal}} = H_{\text{\tiny Applied}} - GM$, where $H$ is either the internal or applied magnetic field, $G$ is the geometric factor resulting from the crystal geometry, and $M$ is the sample magnetization. To a first approximation, the difference between $H_{\text{\tiny Internal}}$ and $H_{\text{\tiny Applied}}$ will be determined wholly by $G$, when the only experimental change is the orientation of the crystal and we assume the magnetization to be the same in each situation. Beleggia \emph{et al.}\cite{beleggiajpd39} proposed a scheme for magnetostatic mapping of shapes to equivalent ellipsoids whose geometric factors can be more easily calculated. Following this work, we can estimate the demagnetization factors of the LiHoF$_{4}$ sample used in this work by a rectangular prism with dimensions $L_{x} = 1$~mm, $L_{y} = 2.1$~mm, and $L_{z} = 5.5$~mm.

Using the Mathematica code provided in the above reference, we find the demagnetization factors $G_{\text{\tiny x}} = G_{\text{\tiny max}} = 7.55$, $G_{\text{\tiny y}} = G_{\text{\tiny min}} = 1.68$, and $G_{\text{\tiny z}} = -0.32$. Thus, the correction for the maximum demagnetization orientation will be approximately 4.5 times larger than in the minimum demagnetization orientation. By repeating the same experiments using these two different sample orientations, we can compare demagnetization corrections. As expected, in all orientations, the observed zero-field transitions are identical and occur at $T_{c}$ = 1.532$\pm$0.005~K.  
\begin{figure}
\begin{center}
\scalebox{0.67}{\includegraphics{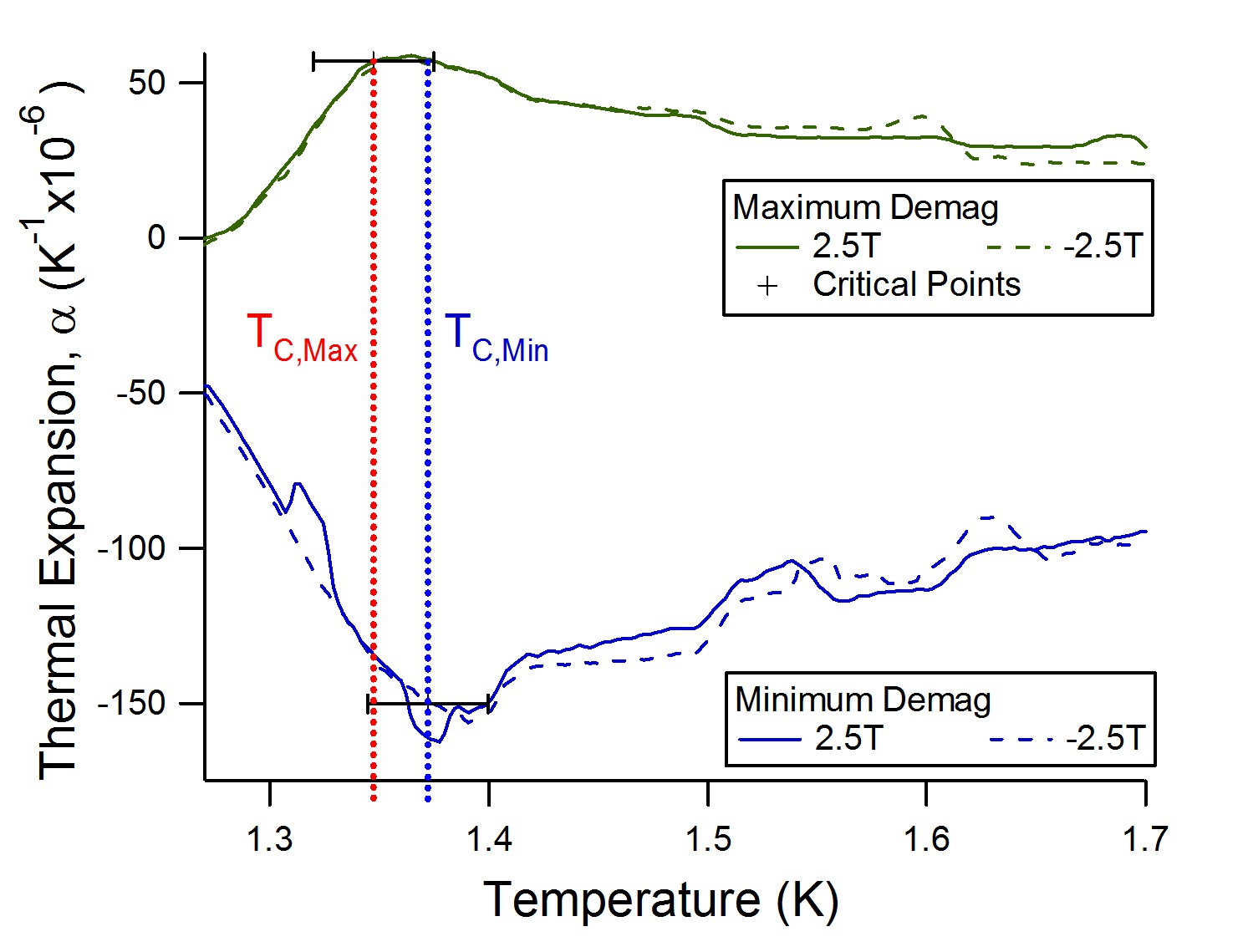}} 
\caption[Effect of torque on $\alpha$ of LiHoF$_{4}$.]{\label{fig:torque}Measurement of $\alpha$ for LiHoF$_{4}$ at $\pm2.5~T$ in the maximum (solid lines) and minimum (dashed lines) demagnetization orientations. Crosses show the calculated critical temperatures. The \textcolor{red}{red} (\textcolor{blue}{blue}) dotted line indicates the critical temperature determined from the \textcolor{red}{maximum} (\textcolor{blue}{minimum}) demagnetization orientation, identical within error.}
\end{center}
\end{figure}
Fig.~\ref{fig:torque} shows the results of such an investigation at $H =$ 2.5~T.  As can be seen, a striking feature of this test is the qualitative difference in the temperature dependence of the thermal expansion between each orientation.  The transition temperature which is extracted in each case is $T_{c}(\text{\small max})$ = 1.35$\pm$0.03~K and $T_{c}(\text{\small min})$ = 1.37$\pm$0.03~K for each orientation as shown by the crosses in the figure.  Consequently, in spite of a difference in the demagnetization factor of almost five, there is only a small difference in the transition temperature which is less than our uncertainty in each value.  This suggests that the demagnetization correction will lead to changes in $T_c$ that are at most within our current error estimates.  

The preceding analysis assumes the sample is single domain.  Calculation of the demagnetisation factor is a far more complicated issue in the presence of domains\cite{Neel,Dunlop}.  Nonetheless, even in the presence of a magnetic domain structure, we would still anticpate an orientation dependence to the demagnetisation factor for this sample and the empirical evidence from this study is that this correction is small.
Theoretical analysis of the domain structure in LiHoF$_4$ has been carried out by Biltmo and Henelius\cite{heneliusepl87}.  We speculate that differences in the domain structure that could occur for each orientation may explain the qualitative orientational difference in the signal that is measured.  A better understanding of the microscopic origin of the thermal expansion coefficient and the behavior of domains in the region of the transition would be required to explain this observation.

\subsection{\label{torque}Torque Effects}
The potential misalignment of a ferromagnetic sample in a transverse magnetic field may lead to a torque on the sample which, if the sample were to rotate or bend, could lead to a signal indistinguishable from thermal expansion or magnetostriction.
In order to verify that any torque applied to LiHoF$_{4}$ was not distorting our ability to measure the phase transition, thermal expansion was measured in positive and negative transverse fields, while in both the minimum and maximum demagnetization orientations.

As shown in Fig.~\ref{fig:torque}, the measured thermal expansion for either field direction is identical.  Consequently we conclude that there is not a significant rotation or bending of the sample due to any putative torque. This is consistent with a small sample magnetic moment due to magnetic domains.

\section{\label{conc}DISCUSSION \& CONCLUSIONS}

\begin{figure}
\begin{center}
\scalebox{0.65}{\includegraphics{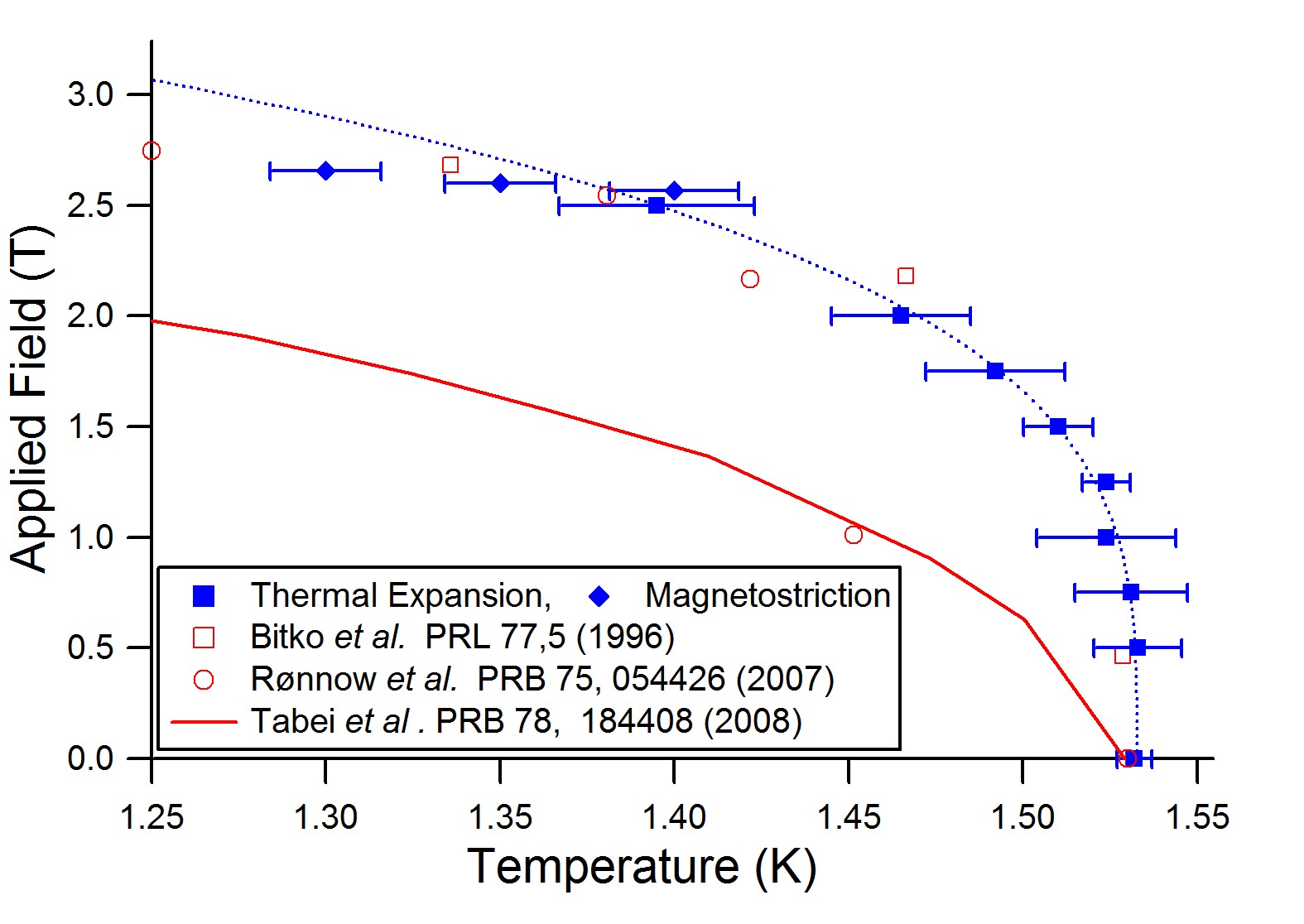}} 
\caption[Phase diagram of LiHoF$_{4}$, including current work.]{\label{fig:PDFinal}Transverse field versus temperature phase diagram of LiHoF$_{4}$, as measured by this study using thermal expansion (squares) and magnetostriction (circles) data.  Also plotted is previous experimental \cite{bitko, ronnowsci308} and theoretical Monte Carlo\cite{tabeiprb78} work. (Dotted line is a guide to the eye.)}
\end{center}
\end{figure}

Combining the ferromagnetic to paramagnetic critical temperatures from thermal expansion measurements with critical fields from magnetostriction measurements produces the phase diagram shown in Fig.~\ref{fig:PDFinal}.  The somewhat broad error bars reflect the broadening of the transition region and consequently the associated difficulty in assigning a transition temperature rather than a loss of precision in our technique.  Also plotted are the critical points from earlier magnetic susceptibility\cite{bitko} and neutron scattering\cite{ronnowsci308} work, and from the latest numerical simulations\cite{tabeiprb78}.  There is excellent agreement between the experimental data from our technique and the magnetic susceptibility, thereby confirming the discrepancy with the current theoretical simulations.  Qualitatively, what appears to occur is that when a transverse field is applied to LiHoF$_4$ in the ferromagnetic state the system is initially impervious to the field.  It is only when the field rises above 1 T, which is approximately 25\% of the field required to completely suppress the ferromagnetism at zero temperature, that the transition temperature starts to be suppressed.  The origin of this imperviousness to the transverse field is unknown and would appear to be an aspect of the physics that is not captured by the transverse field Ising model (Eq. \ref{eq:IsingHam}) used to describe LiHoF$_{4}$ used in numerical simulations\cite{tabeiprb78, ronnowprb75,chakrabortyprb70}. What is now important to assess is whether these results reveal a generic lack of understanding of the impact of quantum fluctuations at finite temperature, or the simpistic nature of the transverse field model as applied to LiHoF$_4$.

One aspect of physics that is specific to LiHoF$_4$ is the effect of magnetoelastic coupling on the antiferromagnetic or quadrupolar exchange parameters as discussed in earlier theoretical studies\cite{ronnowprb75,tabeiprb78}.  These couplings compete with the ferromagnetic coupling and can, in principle, affect the transition temperature.  R\o nnow \emph{et al.} \cite{ronnowprb75} estimated that a strain, $\epsilon_{13}\sim 10^{-4}$ would be required to produce a noticable effect in the transition temperature due to the quadrupolar interactions. Our measurements of magnetostriction are only along the $c$-axis ($\epsilon_{33}$) so we are unable to address this issue directly.   The overall magnitude of the magnetostriction measured in this study is $\Delta l/l \sim 10^{-6}$.  Consequently, unless the magnitude of the in-plane magnetostriction is dramatically different (i.e 2 orders of magnitude larger), it is unlikely that this could lead to a significant restoration of the critical temperature to compensate for the reducing effect of the quantum fluctuations.  Further studies of magnetostriction along the $a$-axis would be required to provide a more complete picture.  In particular, a qualitative change in magnetostrictive behavior between measurements shown here at temperatures very close to $T_c(0)$ and those at lower temperatures would be illuminating.

On a related point, we note that the magnetic field scale of 1 T also signifies a qualitative change in the behaviour of the transition in thermal expansion between the paramagnetic and ferromagnetic state, as seen in Figs.~\ref{fig:lowT},\ref{fig:highT}.  In the absence of a detailed microscopic theory describing the features of the transition, we are not able to assess whether this is a related phenomena or merely coincidental. As already discussed, an experimental investigation of domain wall motion to complement existing theoretical work\cite{heneliusepl87} would be invaluable. Careful analysis of the evolution of the transition in any other technique may also provide additional insight.

Finally, investigation of other Ising systems, for example Ho(OH)$_{3}$ and Dy(OH)$_{3}$, as proposed by Stasiak \emph{et al.} (Ref.~\onlinecite{stasiakprb78}), could determine if this discrepancy between the theoretical model and experimental work is particular to this material, or is a generic feature of the effect of a transverse field on Ising magnetism in real systems.

In conclusion, we have used measurements of thermal expansion and magnetostriction to map the phase diagram of LiHoF$_4$ in a transverse magnetic field at temperatures close to the zero-field classical phase transition where quantum fluctuations are small.  The resultant phase line separating the paramagnetic phase from the ferromagnetic phase is in agreement with earlier experimental studies\cite{bitko} and disagrees with current theoretical results based on the transverse field Ising model\cite{chakrabortyprb70, tabeiprb78}.  The crucial issue now is whether the disagreement pertains to details of the LiHoF$_4$ system or whether it is a generic feature signifying a lack of understanding of the effect of quantum fluctuations at finite temperature.

\section{\label{ack}ACKNOWLEDGEMENTS}
We acknowledge J. Kycia for supplying the LiHoF$_4$ sample used in this study. Thanks go to MJP Gingras, J Kycia, J Quilliam, P Stasiak and A. Tabei for helpful discussion. We would also like to acknowledge the skill of Mike Lang, and the assistance of Harman Vander Heide in the construction of the dilatometer. Financial support was provided by NSERC, the University of Waterloo, and Oxford Instruments PLC.

\bibliography{mybib}
\end{document}